\documentclass[float,aps,10pt]{article}
\usepackage{amsfonts}
\usepackage{amsmath}
\usepackage{amssymb}
\usepackage{epsfig}
\usepackage{graphicx,color}
\usepackage{mathptmx}
\begin{document}

\title{Study of $\bar B^0  \to D^0 \mu^+\mu^-$ Decay in Perturbative QCD Approach}

\author{C.~S. Kim$^{1}$\footnote{Email: cskim@yonsei.ac.kr}~,~~Run-Hui Li$^{1}$
\footnote{Email: lirh@yonsei.ac.kr},~~ Ying Li$^{2}$\footnote{Email: liying@ytu.edu.cn} \\
\small{\it 1.Department of Physics and IPAP, Yonsei University, Seoul
120-479, Korea }\\
\small{\it 2.Department of Physics, Yantai University, Yantai 264-005, China }
}

\maketitle

\begin{abstract}

\noindent Within the perturbative QCD approach and ignoring the contributions of long
distance and subleading penguin loops, we investigate $\bar B^0 \to D^0
\mu^+\mu^-$ decay in the large recoiling kinematic region in the
Standard Model. At the tree level,  $\bar B^0$ decays to $D^0$ by
exchanging a $W$ boson accompanied by a virtual photon emission from
the valence quarks of $\bar B^0$  and $D^0$ meson, then the virtual
photon decays to the lepton pair. Numerically, we find that the
branching ratio decreases rapidly as the $q^2$ increases, and the
branching ratio of $ \bar B^0\to D^0\mu^+\mu^-$ is
$(9.7_{-3.2}^{+4.2})\times 10^{-6}$ in the region $q^2 \in [1,5]~
\mathrm{GeV}^2$. The order of the branching ratio shows a
possibility to study this interesting channel in the current $B$ factories and
the Large Hadron Collider. The precise experimental data will help
us to test the factorization approach and the QCD theory, in general.

\end{abstract}
\newpage

Over the past few years when studying the semileptonic decays of $B$
meson, people always pay much attention on exclusive processes $B
\to (K, K^{*}, \pi, \rho) \ell^+\ell^-$ and inclusive processes $B
\to X_{s,d} \ell^+\ell^-$ as well as similar decay modes, which are
induced by the flavor changing neutral current $b \to s
\ell^+\ell^-$ or $b \to d \ell^+\ell^-$. In these processes, the
leptons are always  generated from either a photon or a $Z$ boson with
loop diagrams, so that these decay processes are
considered as good choices of testing the Standard Model (SM) and
probing possible new physics signals. Recent review in detail
is referred to Refs.
\cite{Antonelli:2009ws,Altmannshofer:2008dz,Buras:2011we}.
 In fact
when we study the decays $B\to (K, K^{*}, \pi, \rho) \ell^+\ell^-$, the weak annihilation
contributions are usually ignored since they are regarded to be
suppressed by $\mathcal{O}(\Lambda_{QCD}/m_B)$ \cite{Beneke:2001at}.
Therefore, we think that it is of urgent interest to explore the
pure annihilation type semileptonic $B$ meson decays, in which
$\mathcal{O}(\Lambda_{QCD}/m_B)$ effects are the main contribution.
Still due to
suppression of $\mathcal{O}(\Lambda_{QCD}/m_B)$,  most of these
decays  have small branching ratios, and cannot be observed in the
current BaBar and Belle experiments. However, for some special
decays, such as $\bar B^0 \to D^0 \mu^+\mu^-$, its branching ratio
can be enhanced by large Wilson coefficients. In this work, we
consider the observables of the decay $\bar B^0 \to D^0 \mu^+\mu^-$
theoretically. Compared with the mass of $B$ meson, both the masses
of muon and electron are very small, so the analysis of $\bar B^0
\to D^0 e^+ e^-$ is almost the same as $\bar B^0 \to D^0
\mu^+\mu^-$.

\begin{figure}[b]
\begin{center}
 \includegraphics[scale=0.4]{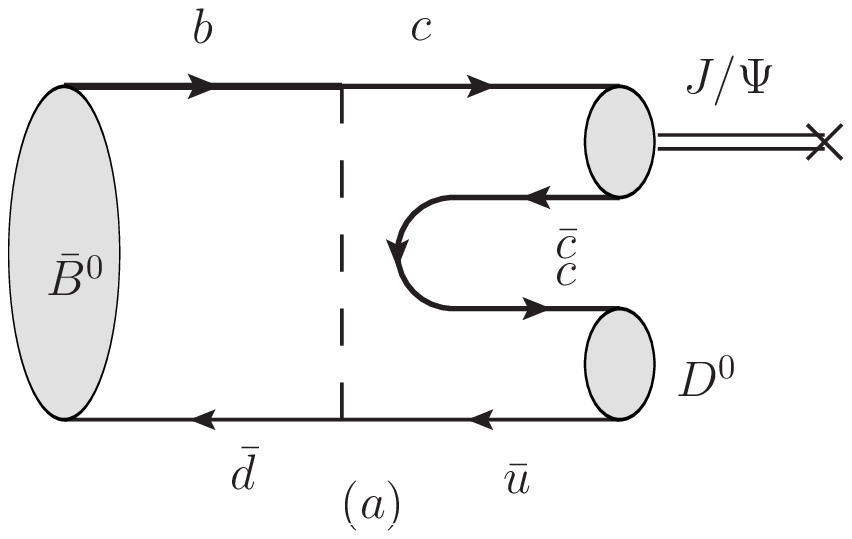}\,\,\,\,\,\,\,\,\,\,~~~~~~~~~~~~~~~~
 \includegraphics[scale=0.4]{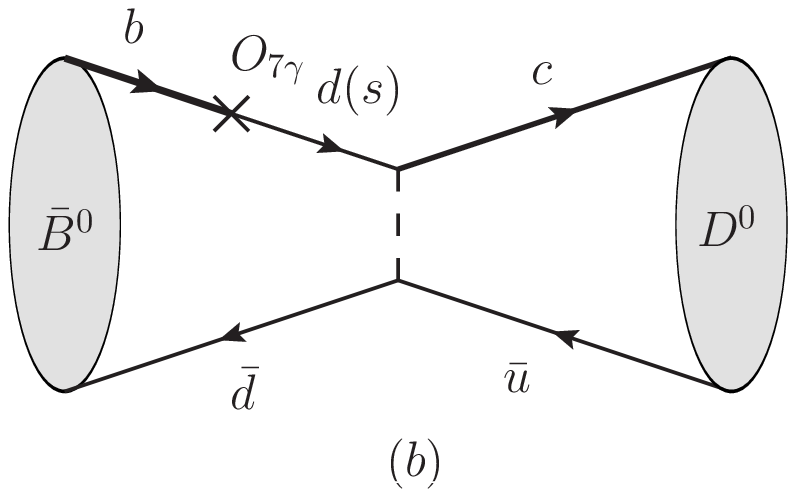}\,\,\,\,\,\,\,\,\,\,~~~~~~~~~~~~~~~~
 \includegraphics[scale=0.4]{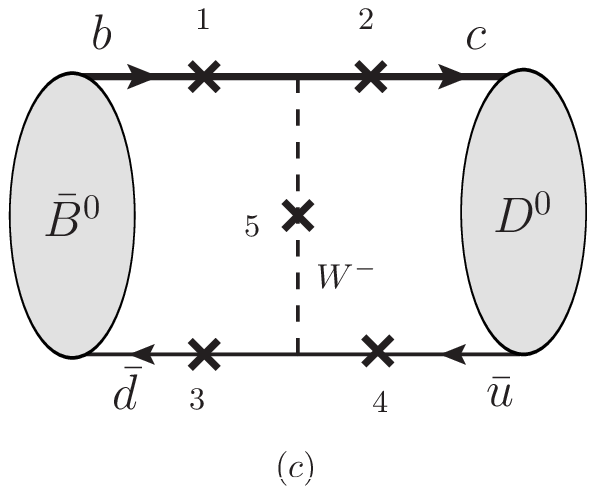}
\caption{The possible diagrams for $B^0 \to D^0 \ell^+\ell^-$, where
the crosses stand for a virtual photon.}
 \label{fig:fd}
 \end{center}
\end{figure}

In the SM for $\bar B^0 \to D^0  \mu^+\mu^-$  the muon pair can be
generated from either a photon or a $Z$ boson, however, the latter
case will be highly suppressed because of the weak coupling and the
large $Z$ mass. Therefore, we only consider the process where the
lepton pair is generated from a virtual photon. In the full theory,
there are three possible contributions to this decay, and we draw
the Feynman diagrams in Fig. \ref{fig:fd}. In the first case, shown
in diagram 1(a), $\bar B^0$ decays to $D^0 + J/\psi$ by exchanging a
$W$ boson and generating $c\bar c$ pair from the vacuum, in which
the $J/\psi$ decays to lepton pair, which is so-called the resonant
contribution. Because the mode $\bar B^0 \to D^0 + J/\psi(\to
\ell^+\ell^-)$ has not been observed yet, we will exclude this part
of contribution, $i.e.$ Fig. 1(a), by carrying out our investigation
in a certain kinematics region, $q^2 \ll m_{J/\psi}^2$. The virtual
photon can also be generated by the penguin operator $O_{7\gamma}^s$
or $O_{7\gamma}^d$, which is shown in diagram 1(b), with the Wilson
coefficient $C_7$.  Since this operator is from the loop suppressed
flavor changing neutral current, the value of $C_7$ is much smaller
than those of the coefficients $C_{1,2}$ of tree operators, and thus
only marginally affecting our numerical estimates.
Therefore, the contribution of diagram 1(b) has been neglected
safely in this work. In diagram 1(c), the $B$ meson decays to a $D$
meson by exchanging a $W$ boson, where the photon can be emitted
from either of the five crosses in diagram. When a photon is emitted
from the $W$ boson, the diagram will be highly suppressed by the two
$W$ propagators and because of the large $W$ mass. Therefore, we
ignore this contribution in our calculation, too. Since this process
happens at the scale ${\cal O}(m_B)$, the highly off-shelled $W$
boson can be integrated out and the effective theory could be used
directly, as shown in Fig. 2.

To make predictions clear, one requires the knowledge of the
matrix element $\langle D \gamma^*|B\rangle$, where the virtual
photon $\gamma^*$ decays to a lepton pair. Although the calculation of
this matrix is not trivial, it has been explored in many approaches,
such as the heavy quark effective theory \cite{Antipin:2006dn}, the heavy
light chiral perturbation theory \cite{Macdonald Sorensen:2006ds}, the
QCD factorization approach \cite{Kivel:2007bq} and the perturbative QCD
(pQCD) approach \cite{Li:2006xe}. Based on $k_T$ factorization,
the pQCD approach \cite{Keum:2000ph1,Keum:2000ph} is one of the
theoretical instruments for handling such exclusive decay modes. The
concept of pQCD is the factorization between soft and hard dynamics.
In this approach, the quark transverse momentum $k_T$ is kept in
order to eliminate the end-point singularity. Because of inclusion
of transverse momenta, double logarithms from the overlap of two
types of infrared divergences, soft and collinear, are generated in
radiative corrections. The resummation of these double logarithms
leads to a Sudakov factor, which suppresses the long-distance
contribution. Though there still exist few controversies
\cite{DescotesGenon:2001hm,Feng:2008zs} on its feasibility, the
predictions based on the pQCD can accommodate experimental data well,
for example, see Ref. \cite{Li:2004ti}. In this work, we will put the
controversies aside and adopt this approach to our analysis.

\begin{figure}[b]
\begin{center}
 \includegraphics[scale=0.7]{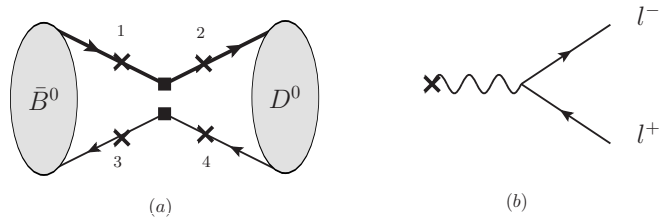}
\caption{Diagram for $B^0 \to D \ell^+\ell^-$ in the effective
theory. The black boxes represent the effective vertex.}
 \label{fig:2}
 \end{center}
\end{figure}

In the SM, the effective Hamiltonian related to decay $\bar B^0 \to
D^0 \ell^+\ell^-$ is given   \cite{Buchalla:1995vs} as:
\begin{eqnarray}
 {\cal
 H}_{\rm{eff}}=\frac{G_F}{\sqrt{2}}V_{cb}V^*_{ud}\left[C_1(\mu)O_1(\mu)+C_2(\mu)O_2(\mu)\right],\label{eq:Heff}
\end{eqnarray}
where $G_F$ is the Fermi constant and $V_{cb}V^*_{ud}$ are the
corresponding CKM matrix elements. $O_1$ and $O_2$ are local
operators, which are defined as:
 \begin{eqnarray}
 O_1&=&(\bar c_{\alpha}b_{\beta})_{V-A}(\bar
 d_{\beta}u_{\alpha})_{V-A}\;\;,\nonumber \\
 O_2&=&(\bar c_{\alpha}b_{\alpha})_{V-A}(\bar
 d_{\beta}u_{\beta})_{V-A}\;\;.
 \end{eqnarray}
Here $\alpha$, $\beta$ are the color indices,  $(\bar
q_1q_2)_{V-A}\equiv\bar q_1\gamma^{\mu}(1-\gamma^5)q_2$, and $C_1$
and $C_2$ are corresponding Wilson coefficients, whose scale evolves
from $m_W$ to the factorization scale $t$. With the Hamiltonian in
Eq.~(\ref{eq:Heff}), we draw the diagram in Fig.~\ref{fig:2}.

Now, we turn to discuss the decay  $\bar B^0 \to D^0 \mu^+\mu^-$ in
certain kinematic regions like $V_{\rm{start}}<q^2<V_{\rm{end}}$,
where $q$ is the momentum of the $\ell^+\ell^-$ pair,
$V_{\rm{start}}$ and $V_{\rm{end}}$ are the boundaries of the
region. To  guarantee our calculation reliable, we should choose the
region where $D$ meson recoils fast and it can be treated on or
nearly on the light cone. In the rest frame of $B$ meson, the
momenta of $B$ and $D$ mesons are defined in the light-cone
coordinate as
\begin{eqnarray}
 p_B=\frac{m_B}{\sqrt{2}}(1,1,\vec{0}_{\perp})\;,\;
 p_D=\frac{m_D}{\sqrt{2}}(\eta+\sqrt{\eta^2-1},\eta-\sqrt{\eta^2-1},\vec{0}_{\perp})\;,\label{eq:momenta}
\end{eqnarray}
with
\begin{eqnarray}
 \frac{m_B^2+m_D^2-V_{\rm{end}}}{2 m_B m_D}< \eta < \frac{m_B^2+m_D^2-V_{\rm{start}}}{2 m_B
 m_D}.\label{eq:eta}
\end{eqnarray}
For the light quarks in $B$ and $D$ mesons, we define their momenta
as
\begin{eqnarray}
k_1=(0,\frac{m_B}{\sqrt{2}}x_1,\vec{k}_{1\perp}),\;  ~~~
k_2=(\frac{\eta+\sqrt{\eta^2-1}}{\sqrt{2}}x_2m_D,0,\vec{k}_{2\perp})\;,
\end{eqnarray}
where $\vec{k}_{\perp}$ stands for the transverse momentum.

For the decay $\bar B^0 \to D^0 \ell^+\ell^-$  the
amplitude will be factorized conventionally to a hadronic part and an
electromagnetic part. To make our expressions simple, we
parameterize the hadronic matrix element with two contracted weak
vertices and one QED vertex as
\begin{eqnarray}
 {T}^{\mu}=\langle D^0| C_i(\mu) O_i(\mu) \;\;
 \frac{(I e_l g)(-I)(I e_q g)\bar q \gamma^{\mu} q}{q^2} |\bar B^0 \rangle
 =f_1(q^2)p_B^{\mu} + f_2(q^2) p_D^{\mu}~,\label{eq:para}
\end{eqnarray}
where $f_1(q^2)$ and $f_2(q^2)$ are form factors, and their
expressions are given by
 \begin{eqnarray}
 f_1(q^2)&=&f_{1,1}(q^2)+f_{1,2}(q^2)+f_{1,3}(q^2)+f_{1,4}(q^2)\;,\;\nonumber\\
 f_2(q^2)&=&f_{2,1}(q^2)+f_{2,2}(q^2)+f_{2,3}(q^2)+f_{2,4}(q^2)\;,
 \end{eqnarray}
in which the second subscripts of $f_{i,j}$ correspond to the numbers
of the crosses in Fig. \ref{fig:2}. Within the perturbative QCD
approach, in the large recoiling region, $f_{i,j}$ could be
calculated at the leading order up to the leading power of
$m_D/m_B$. The detailed expressions are given in Appendix
\ref{sec:formfactors}. Unlike the form factors of the charged
current process $\bar B^0 \to D^-$, $f_1$ and  $f_2$ are complex
numbers, which are caused by the annihilation mechanism. Numerical
results in the region $q^2\in[1\rm{GeV}^2,5\rm{GeV}^2]$ show that
both the real and imaginary parts of $f_1$ are much larger than those
of $f_2$.

With the functions defined above, the amplitude can be expressed as
\begin{eqnarray}
{\cal M}=\frac{G_F}{\sqrt{2}}V_{cb}V^*_{ud}{ T}^{\mu}[\bar
l\gamma_{\mu}l]=f_1(q^2)[\bar l \not p_B l]+f_2(q^2)[\bar l \not p_D
l],
\end{eqnarray}
and
\begin{eqnarray}
|{\cal M}|^2&=& \frac{G_F^2}{2}|V_{cb} V^*_{ud}|^2
\left[|f_1(q^2)|^2 {S}_{11} + |f_2(q^2)|^2 {S}_{22}  +
f_1(q^2)f_2^*(q^2) {S}_{12} +f_1^*(q^2) f_2(q^2){S}_{21} \right]
\end{eqnarray}
with
\begin{eqnarray}
{S}_{11}=Tr[(\not p_1+m_l) \not p_B (\not p_2-m_l) \not p_B],\nonumber\\
{S}_{12}=Tr[(\not p_1+m_l) \not p_B (\not p_2-m_l)\not p_D],\nonumber\\
{S}_{21}=Tr[(\not p_1+m_l) \not p_D (\not p_2-m_l)\not p_B],\nonumber\\
{S}_{22}=Tr[(\not p_1+m_l) \not p_D (\not p_2-m_l) \not p_D].
\end{eqnarray}
In the above functions, $p_1$ and $p_2$ are the momenta of the $l^-$
and $l^+$ leptons respectively, and $m_l$ is the lepton mass. In the
center of mass frame for the lepton pair, we define $p_1^{\prime}$ and
$p_2^{\prime}$ as corresponding momenta of $p_1$ and $p_2$,
\begin{eqnarray}
p_1^{\prime}&=&(\sqrt{q^2}/2,p\sin\theta\cos\phi,p\sin\theta\sin\phi,p\cos\theta),\nonumber\\
p_2^{\prime}&=&(\sqrt{q^2}/2,-p\sin\theta\cos\phi,-p\sin\theta\sin\phi,-p\cos\theta),
\end{eqnarray}
where $p$ is the magnitude of $3$-component momentum and $p^2=q^2/4
-m_l^2$, $\theta$[$\phi$] is the inclination [azimuth] coordinate of
$l^-$. After the Lorentz transformation, one can get the expressions
for $p_1$ and $p_2$ as follows.
\begin{eqnarray}
p_1&=&(\gamma\sqrt{q^2}/2-\gamma\beta p\cos\theta,p\sin\theta\cos\phi,p\sin\theta\sin\phi,-\gamma\beta\sqrt{q^2}/2+\gamma p\cos\theta),\nonumber\\
p_2&=&(\gamma\sqrt{q^2}/2+\gamma\beta
p\cos\theta,-p\sin\theta\cos\phi,-p\sin\theta\sin\phi,-\gamma\beta\sqrt{q^2}/2-\gamma
p\cos\theta),
\end{eqnarray}
where $\beta=\frac{m_D\sqrt{\eta^2-1}}{m_B-m_D\eta}$ and
$\gamma=(1-\beta^2)^{-1/2}$. As a consequence, the expressions for
${S}_{ij}$ with $i,j=1,2$ are given as
\begin{eqnarray}
{S}_{11}&=& m_B^2 \left(4 m_l^2 \cos^2 \theta +q^2 \sin ^2 \theta
\right)\left[ -1+\gamma ^2\left(1+\beta ^2\right)\right]
,\nonumber\\
{S}_{12}&=& m_B m_D  \left(4 m_l^2 \cos ^2 \theta +q^2 \sin^2 \theta
\right)\left[-\eta + \eta\gamma^2 \left(1+\beta ^2\right)  +2 \beta\gamma ^2 \sqrt{\eta ^2-1}\right],\nonumber\\
{S}_{21}&=& {S}_{12} ,\nonumber\\
{S}_{22}&=& m_D^2 \left(4 m_l^2 \cos ^2 \theta  +q^2 \sin ^2 \theta
\right)\left\{-1+\gamma ^2 \left[-1+2 \eta ^2+\beta ^2 \left(2 \eta
^2-1\right)+4 \beta \eta \sqrt{\eta ^2-1}\right]\right\} .
\end{eqnarray}

The most important inputs of the calculation are hadron distribution
amplitudes, named $\phi_B$ and $\phi_D$, which contain the
nonperturbative effects in the mesons under the scale
$\Lambda_{QCD}$. Under the factorization frame, they are universal
quantities and can be constrained from  well measured decay
channels. For the $B$ meson distribution amplitude, we adopt the
model \cite{Keum:2000ph1}:
\begin{eqnarray}
\phi_{B}(x,b)=N_{B}x^{2}(1-x)^{2}\exp \left[ -\frac{1}{2} \left(
\frac{xM_{B}}{\omega _{B}}\right) ^{2} -\frac{\omega
_{B}^{2}b^{2}}{2}\right] \label{bw} \;,
\end{eqnarray}
with the shape parameter $\omega_{B}=0.40\pm0.05$ GeV, which has
been tested in many channels such as $B\to \pi\pi, K\pi$
\cite{Keum:2000ph}. The normalization constant $N_{B}$ is related to
the decay constant $f_{B}=190$ MeV \cite{Keum:2000ph1} by the
normalization condition in Eq. (\ref{eq:normalization}). As for $D$
meson, the distribution amplitude, determined in Ref.
\cite{Li:2008ts} by fitting, is
\begin{eqnarray}\label{Ddis}
\phi_D=\frac{1}{2\sqrt{6}}f_D6x(1-x)\left[1+C_D(1-2x)\right]\exp
\left[ -\frac{\omega^2b^2}{2}\right],
\end{eqnarray}
where $C_D=0.5, \omega=0.1$. Both distribution amplitudes are
normalized as:
\begin{eqnarray}
\int^1_0dx\phi_{M}(x)=\frac{f_{M}}{2\sqrt{2N_C}}, \,\,\, M=B, D
.\label{eq:normalization}
\end{eqnarray}

One can obtain the differential decay width by
\begin{eqnarray}
\frac{d\Gamma}{dq^2 d\cos\theta
 d\phi}=\frac{\sqrt{\lambda}}{1024\pi^4
m_B^3}\sqrt{\frac{q^2-4m_l^2}{q^2}}|{\cal M}|^2,\label{eq:dgamma}
\end{eqnarray}
where $\lambda=(m_B^2+m_D^2-q^2)^2-4m_B^2m_D^2$. Integrating over
the angle variables, we would obtain the $q^2$-dependance of the
decay width as well as the branching ratio. In Eq.
(\ref{eq:dgamma}), the factor $\sqrt{\frac{q^2-4m_l^2}{q^2}}$
ensures that the branching ratio at $q^2=4m_l^2$ vanishes, however,
the $q^2$ appearing in the denominator of the photon propagator
generates a pole-like structure at the small $q^2$ region. Since it
is very difficult for the detector to observe leptons with such a
low energy, we simply subtract the region with very small $q^2$
value. In addition, in order to avoid the pollution from long
distance contributions shown in Fig 1(a), we set the maximum value
of $q^2$ as $5~\mathrm{GeV}^2$.

\begin{figure}[t]
\begin{center}
\includegraphics[scale=0.5]{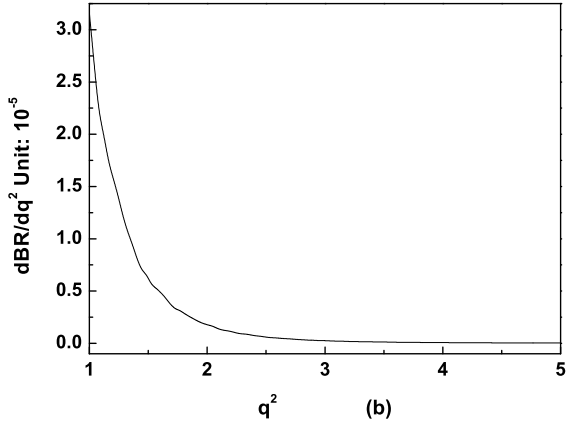}
 \caption{The dependence of the branching ratio of $ \bar B^0\to
D^0\mu^+\mu^-$ with $q^2$,  and $q^2 \in
 [1,5]\rm{GeV}^2 $.}\label{fig:dBR}
\end{center}
\end{figure}

In Fig. \ref{fig:dBR}, we present the behavior of
the branching ratio of this decay mode with $1~{\rm
GeV}^2<q^2<5~{\rm GeV}^2$. From the figure, one can see that the
value of the branching ratio decreases rapidly as the $q^2$
increases: at $q^2=1~\rm{GeV}^2$ the value is $3.2\times10^{-5}$,
and it decreases to $2.8\times10^{-8}$ at $q^2=5~\rm{GeV}^2$.  By
integrating the branching ratio over $q^2$ in the region $[1 ,
5]~\rm{GeV}^2$, we obtain:
\begin{eqnarray}
BR(\bar B^0\to D^0\mu^+\mu^-)=\left( 9.7_{-3.2}^{+4.2} \right)\times
10^{-6},
\end{eqnarray}
where the errors are mainly from $\Lambda_{QCD}$. The errors from
the decay constant are not listed directly, which are proportional
to the square of the decay constants. We here do not discuss the
uncertainties taken by CKM elements, simply because they can be
measured well in other decay channels. Since there  only vector
currents appear in the calculation, there is no forward-backward
asymmetry in this decay mode at the tree level, so any apparent
deviation from zero would be the signal from new physics. The order
of magnitude for branching ratio shows a possibility to study this
channel  in present Belle, BaBar and LHC-b as well as future
Super-$B$ factories. The precise experimental data will help us to
test the factorization approach, and the QCD theory itself in
general. We are pretty sure that future studies on the decays will
come soon from several other theoretical approaches, and the
numerical estimates will be further refined.

Finally, let us summarize our work. Within the pQCD approach, we
studied the exclusive rare decay of $\bar B^0  \to D^0 \mu^+\mu^-$,
which is pure annihilation type decay. Explicitly, we have found
that the branching ratio is $\left( 9.7_{-3.2}^{+4.2} \right)\times
10^{-6}$ and the forward-backward asymmetry is zero at the tree
level. It is clear that such an order of magnitude for branching
ratio could be well measured at the ongoing $B$ factories and
Large Hadron Collider as well as future Super-$B$ factories.

\section*{Acknowledgement}

C.S.K. was supported by the NRF grant funded by the Korea government (MEST)
(No. 2011-0027275) and (No. 2011-0017430).
The work of R.H.L. was supported by the Brain Korea 21 project.
The work of Y.L. was supported in part by the NSFC
(Nos.10805037 and 10625525) and the Natural Science Foundation of
Shandong Province (ZR2010AM036).

\begin{appendix}
\section{Appendix A: Relevant Functions}
\label{sec:formfactors}

The definitions of $f_{i,j}$ used in the text are presented in this
appendix. These functions can be calculated directly within the
perturbative QCD approach:
 {\small
\begin{eqnarray}
 f_{1,1}(q^2)&=&4 e_b \pi\alpha_{\rm{em}} m_D
 f_D \int_0^1 dx_1 \int_0^{\rm{\Lambda_{\rm{QCD}}}}db_1 b_1 a_2(t_1)\exp[-S_B(t_1)]\phi_B(x_1)
 \frac{\sqrt{6}}{(\eta^2-1)q^2\pi}H_0(\sqrt{D_1}b_1)\nonumber\\
 &&\times\left(2 m_D (4\eta^4-5\eta^2-3\eta\sqrt{\eta^2-1}+4\eta^3\sqrt{\eta^2-1}+1)
 -m_B(x_1-2)(2\eta^3+2\eta^2\sqrt{\eta^2-1}-\sqrt{\eta^2-1}-2\eta)\right),\nonumber\\
 f_{2,1}(q^2)&=&4 e_b \pi \alpha_{\rm{em}}
 m_B f_D \int_0^1 dx_1 \int_0^{\rm{\Lambda_{\rm{QCD}}}}db_1 b_1 a_2(t_1)\exp[-S_B(t_1)]\phi_B(x_1)
 \frac{\sqrt{6}}{(\eta^2-1)q^2 \pi}H_0(\sqrt{D_1}b_1)\nonumber\\
 &&\times\left(m_B(x_1-2)(\eta^2+\eta\sqrt{\eta^2-1}-1)
 +2m_D (-2\eta^3-2\eta^2\sqrt{\eta^2-1}+\sqrt{\eta^2-1} +2\eta)\right),\nonumber\\
 f_{1,2}(q^2)&=&4 e_d \pi \alpha_{\rm{em}} m_D
 f_D \int_0^1 dx_1 \int_0^{\rm{\Lambda_{\rm{QCD}}}}db_1 b_1 a_2(t_2)\exp[-S_B(t_2)]\phi_B(x_1)
 \frac{\sqrt{6}}{(\eta^2-1)q^2 }\nonumber\\
 &&\times\left\{
 \begin{array}{l}
 \frac{1}{\pi} K_0(\sqrt{D_2} b_1)\,\,\,\,\,\,\,\mbox{when}\,\, D_2>0
 \\
 \frac{i}{2} H_0(\sqrt{|D_2|} b_1)\,\,\,\mbox{when}\,\, D_2<0
 \end{array}\right\}
 \nonumber\\
 &&\times\left(2 m_D (4\eta^4-5\eta^2-3\eta\sqrt{\eta^2-1}+4\eta^3\sqrt{\eta^2-1}+1)
 +m_B(x_1-2)(2\eta^3+2\eta^2\sqrt{\eta^2-1}-\sqrt{\eta^2-1}-2\eta)\right),\nonumber\\
 f_{2,2}(q^2)&=&-4 e_d \pi \alpha_{\rm{em}}
 m_B f_D \int_0^1 dx_1 \int_0^{\rm{\Lambda_{\rm{QCD}}}}db_1 b_1 a_2(t_2)\exp[-S_B(t_2)]\phi_B(x_1)
 \frac{\sqrt{6}}{(\eta^2-1)q^2}\nonumber\\
 &&\times\left\{
 \begin{array}{l}
 \frac{1}{\pi} K_0(\sqrt{D_2} b_1)\,\,\,\,\,\,\,\mbox{when}\,\, D_2>0
 \\
 \frac{i}{2} H_0(\sqrt{|D_2|} b_1)\,\,\,\mbox{when}\,\, D_2<0
 \end{array}\right\}\nonumber\\
 &&\times\left(m_B(x_1-2)(\eta^2+\eta\sqrt{\eta^2-1}-1)
 +2m_D (2\eta^3+2\eta^2\sqrt{\eta^2-1}-\sqrt{\eta^2-1}
 -2\eta)\right),\nonumber
 \end{eqnarray}
 \begin{eqnarray}
 f_{1,3}(q^2)&=&-4 e_c \pi \alpha_{\rm{em}} m_D
 f_B \int_0^1 dx_2 \int_0^{\rm{\Lambda_{\rm{QCD}}}}db_2 b_2 a_2(t_3)\exp[-S_D(t_3)]\phi_D(x_2)
 \frac{\sqrt{6}}{(\eta^2-1)q^2}\nonumber\\
 &&\times\left\{
 \begin{array}{l}
 \frac{1}{\pi} K_0(\sqrt{D_3} b_2)\,\,\,\,\,\,\,\mbox{when}\,\, D_3>0
 \\
 \frac{i}{2} H_0(\sqrt{|D_3|} b_2)\,\,\,\mbox{when}\,\, D_3<0
 \end{array}\right\}
 \times\left(2 m_B \left(-2\eta^3-2\eta^2\sqrt{\eta^2-1}+\sqrt{\eta^2-1}+2\eta\right) \right.\nonumber\\
 &&\left.+m_D\left(x_2(4\eta^4-5\eta^2-3\eta\sqrt{\eta^2-1}+4\eta^3\sqrt{\eta^2-1}+1)
 -2(\eta^2+\eta\sqrt{\eta^2-1}-1)\right)\right),\nonumber\\
 f_{2,3}(q^2)&=&-4 e_c \pi \alpha_{\rm{em}}
 m_B f_B \int_0^1 dx_2 \int_0^{\rm{\Lambda_{\rm{QCD}}}}db_2 b_2 a_2(t_3)\exp[-S_D(t_3)]\phi_D(x_2)
 \frac{\sqrt{6}}{(\eta^2-1)q^2 }\nonumber\\
 &&\times\left\{
 \begin{array}{l}
 \frac{1}{\pi} K_0(\sqrt{D_3} b_2)\,\,\,\,\,\,\,\mbox{when}\,\, D_3>0
 \\
 \frac{i}{2} H_0(\sqrt{|D_3|} b_2)\,\,\,\mbox{when}\,\, D_3<0
 \end{array}\right\}\nonumber\\
 &&\times\left(2m_B(\eta^2+\eta\sqrt{\eta^2-1}-1)
 +m_D\left(x_2(-2\eta^3-2\eta^2\sqrt{\eta^2-1}+\sqrt{\eta^2-1}+2\eta)+2\sqrt{\eta^2-1}\right)\right),\nonumber\\
 f_{1,4}(q^2)&=&-4 e_u \pi \alpha_{\rm{em}} m_D
 f_B \int_0^1 dx_2 \int_0^{\rm{\Lambda_{\rm{QCD}}}}db_2 b_2 a_2(t_4)\exp[-S_D(t_4)]\phi_D(x_2)
 \frac{i\sqrt{3}}{\sqrt{2}(\eta^2-1)q^2}H_0(\sqrt{D_4}b_2)\nonumber\\
 &&\times\left(2 m_B \left(2\eta^3+2\eta^2\sqrt{\eta^2-1}-\sqrt{\eta^2-1}-2\eta\right)
 +m_D(x_2-2)\left(4\eta^4-5\eta^2-3\eta\sqrt{\eta^2-1}+4\eta^3\sqrt{\eta^2-1}+1\right)\right),\nonumber\\
 f_{2,4}(q^2)&=&-4 e_u \pi \alpha_{\rm{em}}
 m_B f_B \int_0^1 dx_2 \int_0^{\rm{\Lambda_{\rm{QCD}}}}db_2 b_2 a_2(t_4)\exp[-S_D(t_4)]\phi_D(x_2)
 \frac{i\sqrt{3}}{\sqrt{2}(\eta^2-1)q^2}H_0(\sqrt{D_4}b_2)\nonumber\\
 &&\times\left(-2m_B(\eta^2+\eta\sqrt{\eta^2-1}-1)
 -m_D(x_2-2)\left(2\eta^3+2\eta^2\sqrt{\eta^2-1}-\sqrt{\eta^2-1}-2\eta\right)\right),\label{eq:formula}
\end{eqnarray}
 }
where $H_0^{(1)}(z) = J_0(z) + i\, Y_0(z)$, and $J_0, Y_0$ and
$K_0$ are Bessel functions.

The expressions for $D_i$ ($i=1,2,3,4$)
are given as
\begin{eqnarray}
 D_1&=&-m_D^2+m_B^2+m_B m_D x_1 (\eta+\sqrt{\eta^2-1})\;,\nonumber\\
 D_2&=&-m_B^2(1-x_1) - m_D^2 - m_B m_D [-2\eta + x_1(\eta+\sqrt{\eta^2-1})],\nonumber\\
 D_3&=&-m_B^2 + m_D^2 +m_B m_D x_2 (\eta+\sqrt{\eta^2-1}),\nonumber\\
 D_4&=&-m_B^2 - 2m_D^2 +m_B m_D(\eta-\sqrt{\eta^2-1}).
\end{eqnarray}

The hard scale $t$'s in the amplitudes are taken as the largest
energy scale in the hard kernel $H_0$ (or $K_0$):
$t_i={\rm{max}}\left(\sqrt{|D_i|},1/b_j\right)$ with $j=1$ when
$i=1,2$ and $j=2$ when $i=3,4$.
\end{appendix}

Functions, $S_B$ and $S_D$, result from summing both double
logarithms caused by soft gluon corrections and singular ones due to
the renormalization of ultra-violet divergence. $S_{B, D}$
are defined as
\begin{gather}
S_B(t) = s(x_1P_1^+,b_1) +
2 \int_{1/b_1}^t \frac{d\mu'}{\mu'} \gamma_q(\mu'), \\
S_D(t) = s(x_2P_2^+,b_3) + 2 \int_{1/b_2}^t \frac{d\mu'}{\mu'}
\gamma_q(\mu'),
\end{gather}
where $s(Q,b)$, so-called Sudakov factor, is given  in \cite{Keum:2000ph} as
\begin{eqnarray}
  s(Q,b) &=& \int_{1/b}^Q \!\! \frac{d\mu'}{\mu'} \left[
 \left\{ \frac{2}{3}(2 \gamma_E - 1 - \log 2) + C_F \log \frac{Q}{\mu'}
 \right\} \frac{\alpha_s(\mu')}{\pi} \right. \nonumber \\
& &  \left.+ \left\{ \frac{67}{9} - \frac{\pi^2}{3} - \frac{10}{27}
n_f
 + \frac{2}{3} \beta_0 \log \frac{\gamma_E}{2} \right\}
 \left( \frac{\alpha_s(\mu')}{\pi} \right)^2 \log \frac{Q}{\mu'}
 \right],
 \label{eq:SudakovExpress}
\end{eqnarray}
where $\gamma_E=0.57722\cdots$ is Euler constant,
and $\gamma_q = \alpha_s/\pi$ is the quark anomalous dimension.

\end{document}